\documentclass[aps,showpacs,floatfix,twocolumn,superscriptaddress]{revtex4-1}
\usepackage{graphicx}
\usepackage{bm,amsmath}
\usepackage{dcolumn}
\usepackage{amsfonts,amssymb}

\usepackage{bbm, dsfont}
\usepackage{color}

\usepackage{hyperref,cleveref}

\DeclareMathOperator{\re}{Re}
\DeclareMathOperator{\diag}{diag}
\def \P {\hat{\mathcal{P}}}
\def \T {\hat{\mathcal{T}}}

\def \PT {\mathcal{PT}}

\begin{document}

\title{$\mathcal{PT}$-symmetric non-Hermitian Dirac semimetals}

 \author{W. B. Rui}
 %\email[]{w.rui@fkf.mpg.de}
 \affiliation{Max-Planck-Institute for Solid State Research, Heisenbergstrasse 1, D-70569 Stuttgart, Germany}
 
 \author{Moritz M. Hirschmann}
  %\email[]{m.hirschmann@fkf.mpg.de}
 \affiliation{Max-Planck-Institute for Solid State Research, Heisenbergstrasse 1, D-70569 Stuttgart, Germany}

 \author{Andreas P.\ Schnyder}
 %\email[]{a.schnyder@fkf.mpg.de}
 \affiliation{Max-Planck-Institute for Solid State Research, Heisenbergstrasse 1, D-70569 Stuttgart, Germany}
 
 \date{\today}

\begin{abstract}
Parity-time ($\PT$) symmetry plays an important role both in non-Hermitian and topological systems. 
In non-Hermitian systems  $\PT$ symmetry can lead to an entirely real energy spectrum, while 
in topological systems $\PT$ symmetry gives rise  
to stable and protected Dirac points. 
Here, we study a $\PT$-symmetric system which is both non-Hermitian and topological, namely a 
$\PT$-symmetric Dirac semimetal with non-Hermitian perturbations in three dimensions.
We find that, in general, there are only two types of symmetry allowed non-Hermitian perturbations, namely non-Hermitian kinetic potentials, and non-Hermitian anti-commuting potentials.
For both of these non-Hermitian potentials we investigate the band topology for open and periodic boundary conditions, determine the exceptional 
points, and compute the surface states. We find that with periodic boundary conditions, the non-Hermitian kinetic potential leads to exceptional rings, while the non-Hermitian anti-commuting potential generates exceptional spheres, which separate regions with broken $\PT$ symmetry from regions with unbroken  $\PT$ symmetry. 
With open boundary conditions,  the non-Hermitian kinetic potential induces a non-Hermitian skin effect which is localized on both sides of the sample due to symmetry, while the non-Hermitian
anticommuting potential leads to Fermi ribbon surface states. 
\end{abstract}

\maketitle

\section{Introduction}

 Parity-time ($\PT$) symmetry, which inverts both time and spatial coordinates, 
plays an important role in non-Hermitian systems, where it can lead to purely real energy spectra~\cite{bender_pt-symmetric_1999,dorey_spectral_2001,bender_complex_2002,Bender_introduction_2007,Lee_local_2014,konotop_nonlinear_2016}.
This has been demonstrated recently in numerous photonics experiments, where $\PT$ symmetry can 
be readily implemented by balancing radiation gain and loss~\cite{makris_beam_2008,guo_observation_2009,ruter_observation_2010,feng_non-hermitian_2017,el-ganainy_non-hermitian_2018}. 
These experiments have deepened our understanding of $\PT$-symmetric non-Hermitian physics and have revealed
a number of interesting phenomena, such as unidirectional invisibility and single-mode lasing, that could be exploited for applications~\cite{lin_unidirectional_2011,feng_single-mode_2014,goldzak_light_2018,chen_exceptional_2017,hodaei_enhanced_2017}.
Intriguingly, it has been found that large non-Hermitian perturbations can drive the system into a $\PT$-symmetry broken phase, where
the $\PT$ operator and the Hamiltonian no longer have the same eigenstates, even though they commute~\cite{bender_pt-symmetric_1999,feng_non-hermitian_2017}.
As a consequence, the spectrum in the $\PT$-broken phase becomes complex.
The $\PT$-broken phase  is separated from the $\PT$-unbroken phase by exceptional points, i.e.,   degeneracies at which
two or more eigenstates become identical~\cite{konotop_nonlinear_2016,heiss_physics_2012}.

$\PT$-inversion is also an important symmetry in many topological systems. 
In particular, in topological semimetals it can lead to stable and protected band crossings, such as Dirac nodal lines or
Dirac points~\cite{chan_nodal_line_PRB_16,pt_cp_top_semi_metal_PRL_16,zhao_$pt$-symmetric_2017,chiu_classification_2016,Armitage_Weyl_2018}.
We note that for the protection of these band crossing only the combined symmetry of $\mathcal{P}$ and $\mathcal{T}$ is required,
while $\mathcal{P}$ and $\mathcal{T}$ can be broken individually~\cite{tang_dirac_2016,smejkal_electric_2017}.
At these band degeneracies, 
linearly dispersing bands cross each other, leading to   fourfold degeneracies~\cite{liu_discovery_2014,liu_stable_2014,Wang_three_2014,neupane_observation_2014,borisenko_experimental_2014,yang_classification_2014}.  
The topology of these band crossings is characterized by a binary $\mathbbm{Z}_2$ invariant, i.e., either by a quantized $\pi$-Berry phase
or a $\mathbbm{Z}_2$ monopole charge. 
By the bulk-boundary correspondence, these invariants lead to drumhead or Fermi arc surface states.

%%% 
In this paper, we want to study a $\PT$-symmetric system, which is both non-Hermitian and topological, namely the $\PT$-symmetric real Dirac semimetal~\cite{pt_cp_top_semi_metal_PRL_16,zhao_$pt$-symmetric_2017}
with non-Hermitian potentials. The stability of the real Dirac points in such a semimetal is guaranteed by a $\PT$ symmetry that squares to $+1$, i.e., $(\PT)^2=+1$.
By a judicious choice of basis, the Hamiltonian describing this Dirac semimetal can be chosen to be purely real, in which case the $\PT$ operator 
simplifies to the complex conjugation operator $\mathcal{K}$, i.e., $\PT = \mathcal{K}$. Despite early attempts on spinless non-Hermitian Dirac semimetals~\cite{zhen_spawning_2015,okugawa_topological_2019}, in view of recent developments on topological gapless phases~\cite{xu_weyl_2017,zyuzin_flat_2018,Papaj_nodal_2019,lee_anomalous_2016,zhou_observation_2018,shen_topological_2018,gong_topological_2018,yao_edge_2018,kawabata_symmetry_2018,borgnia_non-hermitian_2019,ghatak_new_2019}, several crucial questions have not been addressed in the spinful non-Hermitian Dirac semimetals in three dimensions: How do non-Hermitian potentials deform the real Dirac points? What type of exceptional manifolds~\cite{yang_non-hermitian_2019,carlstrom_knotted_2019,Yoshida_symmetry_2019} do they generate? How does this depend on the boundary conditions?
What type of surface states exist in $\PT$-symmetric non-Hermitian Dirac semimetals, and is there a bulk-boundary correspondence that relates
them to the bulk topology?

In general we find that for the $\PT$-symmetric real Dirac semimetal there exist only two different types of symmetry-allowed non-Hermitian potentials, namely
the non-Hermitian kinetic potential and the non-Hermitian anti-commuting potential.
With periodic boundary conditions (PBCs), the kinetic potential turns the Dirac point into an exceptional ring and drives the semimetal 
into a $\PT$-broken phase with complex spectrum. With open boundary conditions (OBCs), on the other hand, the Dirac semimetal
with non-Hermitian kinetic potential remains in the $\PT$-unbroken phase (except for a small region around the origin). 
Its eigenstates show a non-Hermitian skin effect, where both bulk and surface states are exponentially localized on
the two surfaces of the semimetal. 
 
The non-Hermitian anti-commuting potential, in contrast, deforms the Dirac points of the periodic system into   exceptional spheres. 
These exceptional spheres separate the Brillouin zone into   $\PT$-broken and   $\PT$-unbroken regions. The topology of 
these exceptional spheres is described by a $\mathbb{Z}_2$ monopole charge, similar to the Hermitian case~\cite{zhao_$pt$-symmetric_2017}.
With OBCs, the non-Hermitian anti-commuting potential gives rise
to  Fermi ribbon surface states, i.e.,   two-dimensional regions of surface states that connect the projections
of the exceptional spheres. Remarkably, within this two-dimensional region the surface states have vanishing real energy.

The aforementioned cases of different non-Hermitian potentials and boundary conditions are summarized in Table~\ref{Table:PT-phase}. A common feature is that systems with $\PT$-unbroken phases retain the topological property of the original real $\mathbb{Z}_2$ monopole charge, where the bulk-boundary correspondence can also be established.

The rest of this paper is structured as follows. 
In Sec.~\ref{sec_continuum_model} we present the continuum model of the  $\PT$-symmetric non-Hermitian Dirac semimetal
and discuss the symmetry allowed non-Hermitian potentials. 
In Sec.~\ref{non-Hermitian-kinetic}, we introduce the lattice model of the $\PT$-symmetric Dirac semimetal and discuss the case of $\PT$-symmetric non-Hermitian kinetic potentials. 
In Sec.~\ref{non-Hermitian-anti-commuting}, we discuss $\PT$-symmetric non-Hermitian potentials that anticommute with the Dirac Hamiltonian. 
Finally, we conclude and discuss our findings in Sec.~\ref{conclusion-and-discussion}.

%%%%%%%%%%%%%%%%%%%%%%%%%%%%%%%%%%%%%%%%%%%%%%%%%%%%%%%%%
\begin{table}[tb]
	\centering 
	\begin{tabular}{c|ccc}
		\hline\hline
		non-Hermiticity &  PBC & OBC & consistent \\  \hline
		kinetic  & $\PT$-broken & $\PT$-unbroken & no  \\
		anti-commuting  &  $\PT$-unbroken & $\PT$-unbroken & yes \\
		\hline\hline
	\end{tabular}
	\caption{The phases of the $\PT$-symmetric non-Hermitian Dirac semimetals with different symmetry-preserving potentials under different boundary conditions. The last column indicates whether the PBC system is consistent with the OBC system. }
	\label{Table:PT-phase}
\end{table}
%%%%%%%%%%%%%%%%%%%%%%%%%%%%%%%%%%%%%%%%%%%%%%%%%%%%%%%%%

\section{Continuum model of $\PT$-symmetric non-Hermitian Dirac semimetal}
\label{sec_continuum_model}

The Hermitian continuum model of the real Dirac semimetal that respects $\PT$ symmetry, with $\PT^2=+1$, reads~\cite{pt_cp_top_semi_metal_PRL_16,zhao_$pt$-symmetric_2017},
\begin{equation}\label{Dirac_hal}
\mathcal{H}_\text{Dirac}(\mathbf{k})=k_x\Gamma_1+k_y\Gamma_2+k_z\Gamma_3,
\end{equation} 
with the gamma matrices $\Gamma_1=\sigma_1\otimes\tau_0,\Gamma_2=\sigma_2\otimes\tau_2$ and $\Gamma_3=\sigma_3\otimes\tau_0$. Together with $\Gamma_4=\sigma_2\otimes\tau_1$ and $\Gamma_5=\sigma_2\otimes\tau_3$  these gamma matrices form a matrix representation of the Clifford algebra. Here $\sigma_0$ and $\tau_0$ are two-dimensional identity matrices, and $\sigma_i$ and $\tau_i$ with $i=1,2,3$ are Pauli matrices. 
Hamiltonian~\eqref{Dirac_hal} is purely real and hence symmetric under $\PT = \mathcal{K}$, which amounts to complex conjugation. 
The Dirac point is characterized by a real $\mathbb{Z}_2$ monopole for the $O(N)$ Berry bundle, the reality of which is enforced by the $\PT$ symmetry~\cite{zhao_$pt$-symmetric_2017}. This kind of real Dirac points can be readily realized in various
types of metamaterials, for example, in photonic lattices~\cite{guo_three_2017,Guo_observation_2019}.

The Hermiticity of Eq.~\eqref{Dirac_hal} is broken by the inclusion of $\PT$-symmetric non-Hermitian potentials. These non-Hermitian potentials can be systematically explored by considering the relation between the gamma matrices and $\PT$ symmetry. Notice that under $\PT$ symmetry $\Gamma_1$ to $\Gamma_3$ are even, while $\Gamma_4$ and $\Gamma_5$ are odd. We find that there is a total of three types of $\PT$-symmetric non-Hermitian potentials: $\gamma_0=\Gamma_1\Gamma_2\Gamma_3$, $\gamma_i=\varepsilon_{ijk}\Gamma_j\Gamma_k$, and $\gamma_{4/5}=i\Gamma_{4/5}$, where $\varepsilon_{ijk}$ is the Levi-Civita symbol with $i=1,2,3$. In momentum space, while $\gamma_0$ simply splits a Dirac point to two Weyl points separated by their imaginary energies, the non-Hermitian kinetic potentials $\gamma_i$, which commute with one of the kinetic terms in Eq.~\eqref{Dirac_hal}, and the non-Hermitian anti-commuting potentials $\gamma_{4/5}$, which anticommute with the whole Hamiltonian, induce entirely different phases not seen in Hermitian theory. Thus, in this work we focus on the latter two types of non-Hermitian potentials. Since spectra and states in systems with periodic boundary conditions (PBCs) may be drastically different from those with open boundary conditions (OBCs), it is necessary to consider the $\PT$ phases of systems with PBCs and OBCs separately~\cite{kunst_non-hermitian_2018,xiong_why_2018,Kunst_biothogonal_2018,Yokomizo_non-Bloch_2019}.

%%%%%%%%%%%%%%%%%%%%%%%%%%%%%%%%%%%%%%%%%%%%%%%%%%%%%%%%%
\begin{figure}[t]
	\centering
	\includegraphics[width=0.9\linewidth]{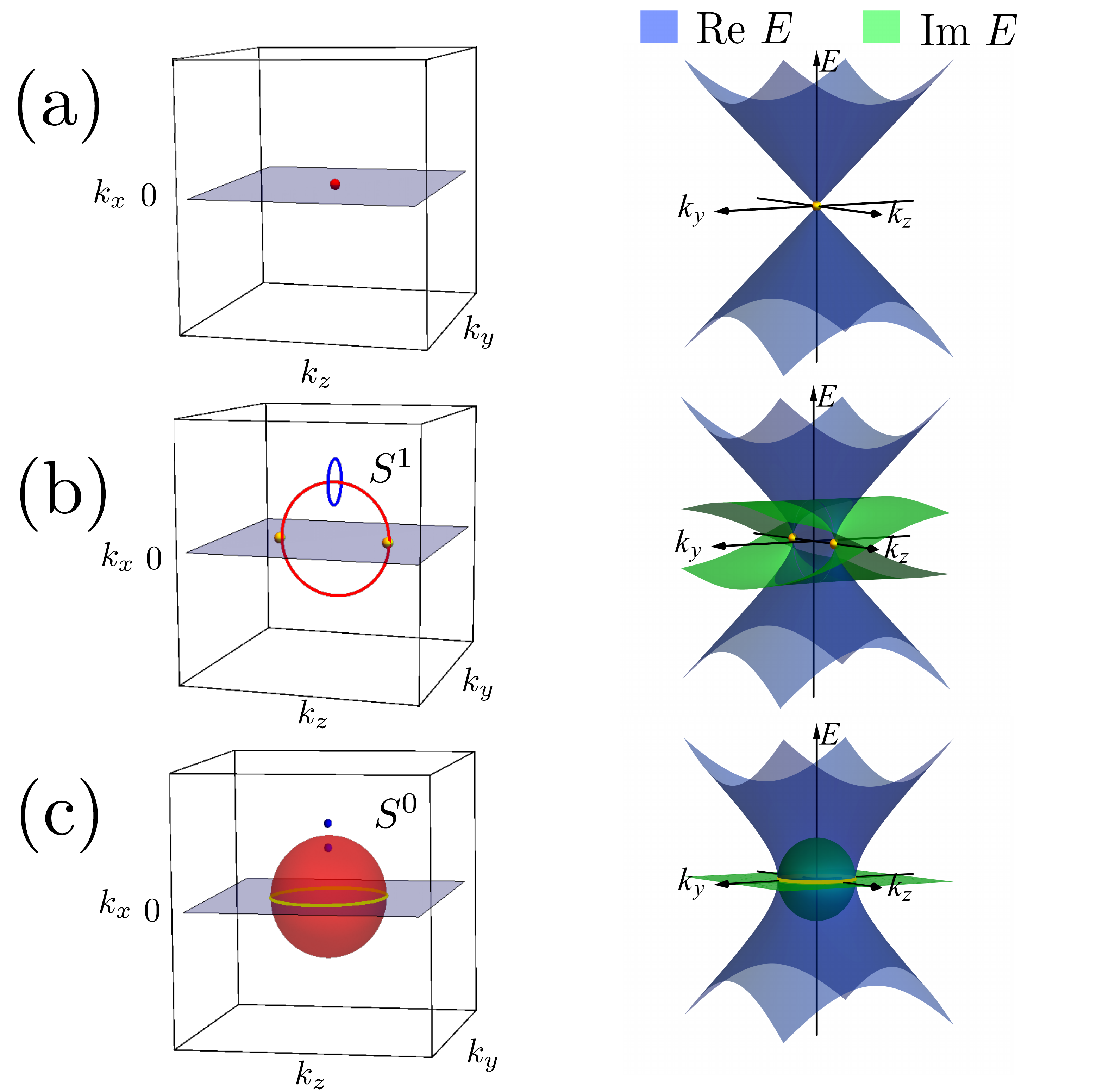}
	\caption{ (a) The left panel shows the Dirac point in the three-dimensional Brillouin zone described by Eq.~\eqref{Dirac_hal}. The energy spectrum on the gray plane ($k_x=0$) against $k_y$ and $k_z$ is shown in the right panel.
    (b) The left panel shows the exceptional ring, which forms after introducing  a $\PT$-symmetric non-Hermitian kinetic potential $\gamma_i$. The corresponding spectrum on the gray plane ($k_x=0$) is shown in the right panel. (c) The left panel shows the case for the $\PT$-symmetric non-Hermitian anti-commuting potential $\gamma_{4/5}$, which induces an exceptional sphere. The corresponding energy spectrum on the gray plane, where the yellow points indicate the band crossings, is shown in the right panel.}
	\label{FIG:exceptional-manifold}
\end{figure}
%%%%%%%%%%%%%%%%%%%%%%%%%%%%%%%%%%%%%%%%%%%%%%%%%%%%%%%%%

\section{$\PT$-symmetric non-Hermitian kinetic potentials}
\label{non-Hermitian-kinetic}

In this section we discuss  $\PT$-symmetric Dirac semimetals perturbed by non-Hermitian kinetic potentials with PBCs.
For this purpose we first construct a minimum lattice model for the Dirac points described by Eq.~\eqref{Dirac_hal}. Similar to Weyl semimetals, Dirac points of $\mathcal{PT}$ symmetric systems appear only in pairs on a lattice, which follows from the Nielsen-Ninomiya theorem~\cite{zhao_$pt$-symmetric_2017}. Thus the minimal lattice model contains at least a pair of Dirac points in the three-dimensional Brillouin zone, which can be constructed as,
\begin{multline}\label{Dirac-lattice}
\mathcal{H}_0(\mathbf{k})=\sin k_x\Gamma_1+\sin k_y\Gamma_2\\+(M-\cos k_x-\cos k_y-\cos k_z)\Gamma_3.
\end{multline}
For $1<M<3$, the two Dirac points are located at $(0,0,\pm k_c)$ with $k_c=\arccos(M-2)$. In the continuum limit with $k_x, k_y\rightarrow 0$ and $k_z\rightarrow \pm k_c$, the lattice model reproduces the form of the continuum model in Eq.~\eqref{Dirac_hal}, with the spectrum around the Dirac point shown in Fig.~\ref{FIG:exceptional-manifold}(a). In a slab geometry with OBCs, the model exhibits Fermi arc surface states that are attached to the two Dirac points in the bulk. 

The lattice model of the $\PT$-symmetric non-Hermitian Dirac semimetals can be simply constructed by including symmetry-preserving non-Hermitian kinetic potentials $\lambda \gamma_i$ as, 
\begin{equation}\label{kinetic-model}
\mathcal{H}_\text{kin}(\mathbf{k})=\mathcal{H}_0(\mathbf{k})+\lambda\gamma_i,
\end{equation}
with $\lambda$ a real quantity. With PBCs, as shown in Fig.~\ref{FIG:exceptional-manifold}(b), Dirac points become exceptional rings under these non-Hermitian potentials. The structure of exceptional rings becomes clearer if we perform a unitary transformation with $U=\exp(i \sigma_0\otimes\tau_1 \pi/4)$ on the Hamiltonian, which yields,
\begin{equation}\label{kinetic-model-block}
U^{-1}\mathcal{H}_\text{kin}(\mathbf{k}) U=\begin{pmatrix}
H_\text{Weyl-ring}(\mathbf{k}) & 0  \\ 
0& H_\text{Weyl-ring}^*(\mathbf{k}) 
\end{pmatrix},
\end{equation}
where $H_\text{Weyl-ring}=\sin k_x\sigma_1+\sin k_y\sigma_2+(M-\cos k_x-\cos k_y-\cos k_z)\sigma_3+i\lambda\sigma_i$. Actually, $H_\text{Weyl-ring}$ is a Hamiltonian describing Weyl exceptional rings, which have been theoretically studied~\cite{xu_weyl_2017,zyuzin_flat_2018} and experimentally realized~\cite{cerjan_experimental_2019}. The two Weyl exceptional rings in $\mathcal{H}_\text{kin}(\mathbf{k})$ can be directly distinguished by their imaginary energies. The energy spectrum of $\mathcal{H}_\text{kin}(\mathbf{k})$ is complex in general, which means the system is in the $\PT$-broken phase. 

A prominent feature of this system is that the band crossings form rings of exceptional points, which possess a non-trivial complex structure~(see discussion in Appendix~\ref{jordan-normal-form}). Associated with the exceptional rings, a topological invariant can be defined. As shown in Fig.~\ref{FIG:exceptional-manifold}(b), we can enclose the exceptional ring with a blue circle $S^1$, and the topological invariant for the $m$-th band with energy $E_m$ is defined as
$\nu_{m}(S^1)=\frac{1}{2\pi}\oint_{S^1} \nabla_{\mathbf{k}} \arg E_m(\mathbf{k})$,
which reflects the winding of the energy eigenvalue on the complex energy plane~\cite{shen_topological_2018,gong_topological_2018}. Here $\arg$ denotes the argument of a complex number. The obtained topological invariant for the occupied band belongs to nontrivial elements of the group $\mathbb{Z}/2$, as discussed in Appendix~\ref{jordan-normal-form}.

We conclude that  $\PT$-symmetric Dirac semimetals with
non-Hermitian kinetic potentials and PBCs are always in the $\PT$-broken phase and the Dirac points become exceptional rings. However, it has been shown in various works~\cite{yao_edge_2018,kunst_non-hermitian_2018,lee_anomalous_2016} that the spectra and states can be quite different in non-Hermitian systems with PBCs and OBCs. It is possible that for the non-Hermitian kinetic potentials, the OBC system might not correspond to the $\PT$-broken PBC system, and instead it might be in a different $\PT$ phase. In the following, we use the transfer matrix method to characterize this difference.

\subsection{Discrepancy between PBC and OBC systems}
In non-Hermitian lattice systems the correspondence between PBC and OBC systems does not always hold, as their eigenstates and spectra can differ from each other. This kind of discrepancy becomes especially crucial for topological phases, because it leads to the failure of bulk-boundary correspondence. Such discrepancy can be qualitatively captured by the transfer matrix $T$ as it is capable of describing systems with open and periodic boundary conditions~\cite{kunst_non-hermitian_2018}. The criterion for the PBC and OBC systems to be consistent with each other is that the transfer matrix is unimodular, i.e., $|\det T|=1$.

For our discussion, we choose the $\PT$-symmetric non-Hermitian kinetic potential as $\gamma_2=\Gamma_3\Gamma_1$, and take OBCs in the $y$ direction.
The transfer matrix for propagation along the $y$ direction, according to Eq.~(7) in Ref.~\cite{kunst_non-hermitian_2018}, can be constructed by identifying the coefficients of $e^{ik_y}$ and $1$ in the Bloch Hamiltonian of Eq.~\eqref{kinetic-model},
\begin{equation}\label{kinetic-transfer-hal}
\mathcal{H}_\text{kin}(\mathbf{k})=J(\tilde{\mathbf{k}})e^{ik_y}+M(\tilde{\mathbf{k}})+J^\dagger (\tilde{\mathbf{k}})e^{-ik_y},
\end{equation}
with $\tilde{\mathbf{k}}=(k_x,k_z)$. The coefficients are found to be
\begin{eqnarray}\label{kinetic-j-m}
&J(\tilde{\mathbf{k}})=\frac{1}{2i}\Gamma_2-\frac{1}{2}\Gamma_3, \\
&M(\tilde{\mathbf{k}})=\sin k_x\Gamma_1+M_{\tilde{\mathbf{k}}}\Gamma_3+\lambda\gamma_2,
\end{eqnarray}
where $M_{\tilde{\mathbf{k}}}=M-\cos k_x-\cos k_z$. Here $J(\tilde{\mathbf{k}})$ and $J^\dagger(\tilde{\mathbf{k}})$ are hopping matrices between two neighboring sites and $M(\tilde{\mathbf{k}})$ is the on-site matrix. The expansion coefficients $\Phi_{n}$ of states at neighboring sites along the $y$ direction are thereby connected by these coefficients, and can be expressed in the transfer matrix equation form $\Phi_{n+1}=T\Phi_n$. The transfer matrix $T$ is obtained in terms of the on-site Green's function 
$\mathcal{G}=[E \mathbbm{1} - M(\tilde{\mathbf{k}})]^{-1}$ and the singular value decomposition of $J(\tilde{\mathbf{k}})=V \Xi W^{\dag}$ as~\cite{kunst_non-hermitian_2018}
\begin{equation}\label{transfer-matrix}
T=\begin{pmatrix}
\Xi^{-1} \mathcal{G}_{vw}^{-1} & -\Xi^{-1} \mathcal{G}_{vw}^{-1}\mathcal{G}_{ww} \Xi \\ 
\mathcal{G}_{vv} \mathcal{G}_{vw}^{-1} & (\mathcal{G}_{wv}-\mathcal{G}_{vv} \mathcal{G}_{vw}^{-1}\mathcal{G}_{ww})\Xi
\end{pmatrix},
\end{equation}
where $\mathcal{G}_{AB}=B^\dagger \mathcal{G}A$ with $A,B\in\{V,W\}$. 

A straightforward calculation shows that the transfer matrix is unimodular, i.e., its determinant is 1. However, from numerical calculations, the OBC spectrum is found to be drastically different from the PBC spectrum, which indicates that the transfer matrix is not unimodular. These two seemingly contradictory results can be resolved by noticing that the system can be block-diagonalized as shown by Eq.~\eqref{kinetic-model-block}. Thus, the transfer matrix can be brought into a block diagonal form, which reads
\begin{equation}
T=\begin{pmatrix}
T_+ &  \\ 
& T_-
\end{pmatrix},
\end{equation}
where
\begin{equation}
T_{\pm}=\frac{1}{M_{\tilde{\mathbf{k}}}\mp \lambda}  \begin{pmatrix}
	\sin^2k_x+M_{\tilde{\mathbf{k}}}^2-\lambda^2-E^2 & \pm \sin k_x+E \\ 
	\pm \sin k_x-E & 1
\end{pmatrix}.
\end{equation}
The determinants of $T_{\pm}$ are,
\begin{equation}
\det T_+=\frac{M_{\tilde{\mathbf{k}}}+\lambda}{M_{\tilde{\mathbf{k}}}-\lambda}, \quad \det T_-=\frac{M_{\tilde{\mathbf{k}}}-\lambda}{M_{\tilde{\mathbf{k}}}+\lambda}.
\end{equation}

In the $\PT$-symmetric non-Hermitian systems with $\lambda\neq0$, though $\det T=\det T_+\det T_{-}=1$ for the entire transfer matrix, individually, neither $\det T_+$ nor $\det T_-$ is $1$. This means there is a marked difference between systems with PBCs and OBCs, and the bulk exceptional rings cannot predict the topological boundary states in the corresponding finite system. Therefore, the OBC system could be in a different $\PT$ phase, which requires a separate investigation.

\subsection{Reflection-symmetric non-Hermitian skin effect}

%%%%%%%%%%%%%%%%%%%%%%%%%%%%%%%%%%%%%%%%%%%%%%%%%%%%%%%%%
\begin{figure}[t]
	\centering
	\includegraphics[width=0.9\linewidth]{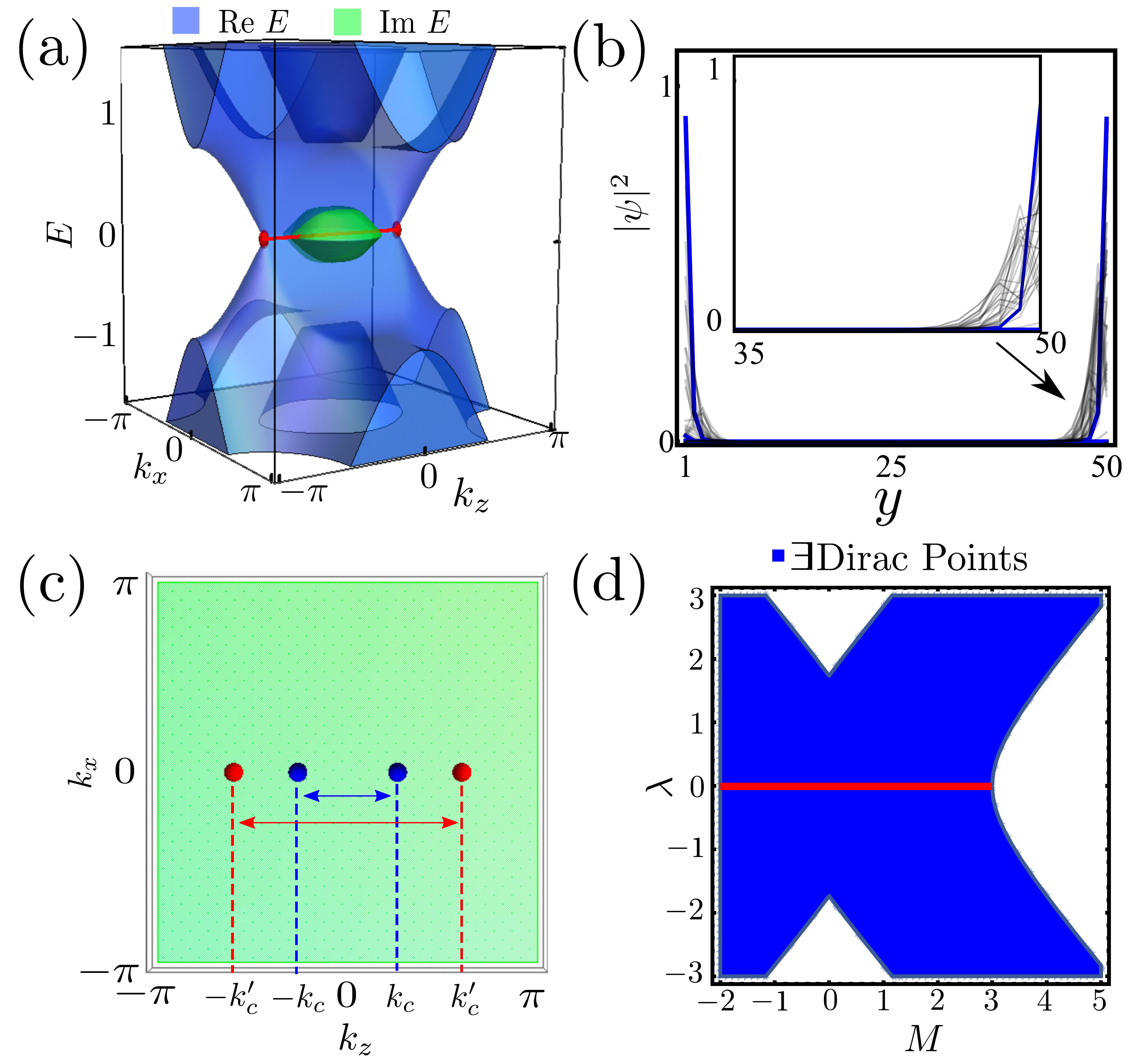}
	\caption{ (a) The OBC spectrum for the $\PT$-symmetric  Dirac semimetal with a non-Hermitian kinetic potential. (b) Wavefunction profile of the OBC system.
    Blue curves are surface states, while black curves are representative bulk states. The parameters are set to be $M=2.0$, $\lambda=0.3$, and $N_y=50$. (c) The location of the Dirac points  (red) in the transformed Hamiltonian of Eq.~\eqref{kinetic-bloch-transformed} and the centers of the exceptional rings (blue) in the original Hamiltonian of Eq.~\eqref{kinetic-model}. (d) Topological phase diagram as a function of mass $M$ and non-Hermitian perturbation $\lambda$.
The blue area denotes the region for which Dirac points exist. The red line corresponds to the Hermitian case.}
	\label{FIG:kinetic}
\end{figure}
%%%%%%%%%%%%%%%%%%%%%%%%%%%%%%%%%%%%%%%%%%%%%%%%%%%%%%%%%

Taking OBCs in the $y$ direction with $N_y$ sites, and PBCs in the $x$ and $z$ directions, the OBC Hamiltonian corresponding to Eq.~\eqref{kinetic-model} reads 
\begin{multline}\label{kinetic-obc}
\mathcal{H}_\text{kin}(\tilde{\mathbf{k}})=
\frac{1}{2i}\Gamma_2\otimes(\widehat{S}-\widehat{S}^\dagger)-\frac{1}{2}\Gamma_3\otimes(\widehat{S}+\widehat{S}^\dagger)
\\ +\left(\sin k_x\Gamma_1+ M_{\tilde{\mathbf{k}}}\Gamma_3+\lambda\gamma_2\right)\otimes\mathbbm{1}_{N_y},
\end{multline}
where $\widehat{S}_{ij}=\delta_{i,j+1}$ is the right-translation operator in the $y$ direction. 

As shown in Fig.~\ref{FIG:kinetic}(a), the spectrum of the OBC system is complex in only a small region near the origin, and outside this region, the system exhibits a $\PT$-unbroken phase with purely real spectrum. The corresponding wavefunction profiles of the OBC system are plotted in Fig.~\ref{FIG:kinetic}(b), which shows that both bulk and boundary modes are localized at two opposite boundaries. Indeed, as predicted by the transfer matrix method, these phenomena in the OBC system are in stark contrast to the PBC system with completely $\PT$-broken phase. 
Thus, it is necessary to work directly with the OBC Hamiltonian $\mathcal{H}_\text{kin}(\tilde{\mathbf{k}})$.

Notice that the reality of the spectrum is ensured in the $\PT$-unbroken phase. It has been shown that a non-unitary similarity transformation can be found to convert a Hamiltonian with real spectrum to be Hermitian~\cite{rui_classification_2019}. For the real space Hamiltonian $\mathcal{H}_\text{kin}(\tilde{\mathbf{k}})$ the similarity transformation is found to be $\mathcal{H}_\text{kin}'(\tilde{\mathbf{k}})= V_{\text{kin}}^{-1} \mathcal{H}_\text{kin}(\tilde{\mathbf{k}})V_{\text{kin}}$, which yields
\begin{multline}\label{kinetic-obc-transformed}
\mathcal{H}_\text{kin}'(\tilde{\mathbf{k}})=\frac{1}{2i}\Gamma_2\otimes(S-S^\dagger)-\frac{1}{2}\Gamma_3\otimes(S+S^\dagger)\\
+\left(\sin k_x\Gamma_1+\sqrt{M_{\tilde{\mathbf{k}}}^2-\lambda^2} \Gamma_3\right)\otimes \mathbbm{1}_{N_y},
\end{multline}
where the similarity transformation operator is $V_{\text{kin}}=\diag\left(\beta^1,\beta^2,\cdots,\beta^y,\cdots\beta^{N_y}\right)$, with $1<y<N_y$ the site index. Here, $\beta^y$ can be brought into block diagonal form by the unitary transformation
\begin{equation}\label{similarity-component}
U^{-1} \beta^y U=\begin{pmatrix}
\alpha^{-y}\rho &  \\ 
& \alpha^y\rho
\end{pmatrix} ,
\end{equation}
where $\rho=(\alpha+1)\sigma_0+(\alpha-1)\sigma_1$ and $\alpha=[(M_{\tilde{\mathbf{k}}}-\lambda)/(M_{\tilde{\mathbf{k}}}+\lambda)]^{1/2}$. Here the unitary operator is $U=\exp(i \sigma_0\otimes\tau_1 \pi/4)$.

In non-Hermitian topological phases, the localization of all eigenstates at the boundary is called the non-Hermitian skin effect~\cite{yao_edge_2018,Kawabata_anomalous_2018,Lee_anatomy_2019}. In previous studies, the localization is found to be on a single boundary.
In our system, the two boundaries are symmetric as the reflection symmetry is preserved in the $y$ direction, $\hat{R}_y \mathcal{H}_\text{kin}(\tilde{\mathbf{k}},-k_y)\hat{R}_y^{-1}=\mathcal{H}_\text{kin}(\tilde{\mathbf{k}},k_y)$ with $\hat{R}_y=\sigma_0\otimes\tau_1$. Hence, the non-Hermitian skin effect manifests itself as the localization of both surface and bulk states at the two boundaries, as shown in Fig.~\ref{FIG:kinetic}(b).
The skin effect can be understood intuitively from the non-unitary similarity transformation $V_\text{kin}$. Let $|\Phi' \rangle$ be the eigenstates of the transformed Hermitian Hamiltonian $\mathcal{H}_\text{kin}'(\tilde{\mathbf{k}})$. Then the eigenstates of the original non-Hermitian Hamiltonian are $|\Phi \rangle=V_\text{kin}|\Phi' \rangle$. According to the explicit form of $V_\text{kin}$ [see Eq.~\eqref{similarity-component}], half components of the states carry a localization factor of $\alpha^{-y}$, and the other half  of $\alpha^{y}$. Therefore, the eigenstates are evenly localized at two opposite boundaries, resulting in a reflection-symmetric non-Hermitian skin effect. Different from previous studies~\cite{yao_edge_2018,Kawabata_anomalous_2018,Lee_anatomy_2019}, the non-Hermitian skin effect discussed here has an intimate relation with symmetry, and locates on both sides of the sample.

\subsection{Dirac points and Fermi arc surface states}

In this subsection we focus on the topological properties of the $\PT$-symmetric Dirac semimetal with
non-Hermitian kinetic potential.
As the OBC system does not correspond to the PBC system, the topological surface states are thereby not in accordance with the bulk exceptional rings. To reveal the topological features of the OBC system, we start with the Fourier transformed version of Hamiltonian~\eqref{kinetic-obc-transformed}. After a Fourier transformation 
in the $y$ direction, this Hamiltonian corresponds to the Bloch Hamiltonian of 
\begin{multline}\label{kinetic-bloch-transformed}
\mathcal{H}_\text{kin}'(\mathbf{k})=\sin k_x\Gamma_1+\sin k_y \Gamma_2\\
+\left(\sqrt{(M-\cos k_x-\cos k_z)^2-\lambda^2} -\cos k_y\right) \Gamma_3.
\end{multline}

Remarkably, the above Hamiltonian exhibits real Dirac points in the Brillouin zone in the Hermitian region, with a Taylor expansion around $(0,0,\pm k_c')$, with $k_c'=\arccos(M-1-\sqrt{\lambda^2+1})$. The Hamiltonian in Eq.~\eqref{kinetic-bloch-transformed} becomes,
\begin{equation}
\mathcal{H}_\text{eff}'(\mathbf{k})= k_x\Gamma_1+  k_y\Gamma_2\pm\gamma_{k_c'}(k_z\mp k_c'),
\end{equation}
with the coefficient $\gamma_{k_c'}= \sqrt{1+\lambda^2}\sqrt{1-\cos^2k_c'}$. 
Clearly, this is a Hermitian Hamiltonian for the real Dirac point described in Eq.~\eqref{Dirac_hal}, for $M$ in the region of $\sqrt{\lambda^2+1}<M<2+\sqrt{\lambda^2+1}$. Notably, the two Dirac points in Eq.~\eqref{kinetic-bloch-transformed} are now relocated to $\mathbf{k}_c'=(0,0,\pm \arccos(M-1-\sqrt{\lambda^2+1}))$, in contrast to the centers of the exceptional rings in Eq.~\eqref{kinetic-model} at $\mathbf{k}_c=(0,0,\pm \arccos(M-2))$. We find that the distance between two Dirac points is elongated compared to that between the centers of the two exceptional rings, as shown by  Fig.~\ref{FIG:kinetic}(c). As highlighted by the red line in Fig.~\ref{FIG:kinetic}(a), the numerically obtained Fermi arc surface states, with purely real spectra, are found to be attached exactly to two red Dirac points in the transformed Bloch Hamiltonian in Eq.~\eqref{kinetic-bloch-transformed}, instead of the centers of the exceptional rings in Eq.~\eqref{kinetic-model}.

It is the main result of this section that with OBCs, in the $\PT$-unbroken phase, the $\PT$-symmetric non-Hermitian Dirac semimetal is equivalent to a Hermitian Dirac semimetal with real Dirac points. From the transformed Hamiltonian $\mathcal{H}_\text{kin}'(\mathbf{k})$, the bulk-boundary correspondence can be recovered successfully. In Fig.~\ref{FIG:kinetic}(d), we plot the region in blue where Dirac points exist in parameter space. Compared with the Hermitian case (red line with $\lambda=0$), the region with Dirac points has been largely expanded. It is verified directly from the non-Hermitian OBC Hamiltonian in Eq.~\eqref{kinetic-obc} that the topological surface states exist in the corresponding region.

\section{$\PT$-symmetric non-Hermitian anti-commuting potentials}
\label{non-Hermitian-anti-commuting}

Now we turn to Dirac semimetals with $\PT$-symmetric non-Hermitian anti-commuting potentials in this section. Here we choose $\gamma_4$ for our discussion. With PBCs, the lattice Hamiltonian in momentum space reads,
\begin{multline}\label{anti-commute-model}
\mathcal{H}_\text{ant}(\mathbf{k})=\sin k_x\Gamma_1+\sin k_y\Gamma_2\\+(M-\cos k_x-\cos k_y-\cos k_z)\Gamma_3+\lambda \gamma_4.
\end{multline}
As shown in Fig.~\ref{FIG:exceptional-manifold}(c), this non-Hermitian potential deforms the two Dirac points into two exceptional spheres. We focus on a single exceptional sphere, which can be described by the continuum model
\begin{equation}
\mathcal{H}_\text{sphere}(\mathbf{k})=k_x\Gamma_1+k_y\Gamma_2+k_z\Gamma_3+\lambda\gamma_4.
\end{equation}
The energy eigenvalues are $E_{\text{sphere},\pm}=\pm (k_x^2+k_y^2+k_z^2-\lambda^2)^{1/2}$. Clearly, the band crossing happens at $k_x^2+k_y^2+k_z^2=\lambda^2$, which describes a sphere. Inside the sphere, the eigenvalues form complex conjugate pairs, and the system is in the $\PT$-broken phase. However, outside the sphere, the system is in the $\PT$-unbroken phase with purely real spectrum. Notably, in the $\PT$-unbroken phase the double degeneracy of bands is still preserved, as it appears in the Hermitian case. We refer to the sphere as an exceptional sphere as it is composed of exceptional points. The complex structure of the exceptional sphere, as discussed in Appendix~\ref{jordan-normal-form}, is associated with a topological invariant defined as follows.

The spatial co-dimension of the exceptional sphere in three dimensions is zero, so that a zero-dimensional sphere $S^0$, which consists of two points of $\mathbf{k}_1$ inside and $\mathbf{k}_2$ outside the sphere, can be chosen to enclose the exceptional sphere. $S^0$ is shown by two blue dots in Fig.~\ref{FIG:exceptional-manifold}(c). Similar to the winding of the energy eigenvalues in the previous section, the topological charge on $S^0$ can be defined as 
$\nu_m(S^0)=\frac{1}{2\pi} (\arg E_{m,\mathbf{k}_1}-\arg E_{m,\mathbf{k}_2})$,
where $E_m$ is the energy of the $m$-th band. Notice that although the topological invariant adopted in Refs.~\cite{okugawa_topological_2019,kawabata_classification_2019} for the $\mathcal{PT}$-symmetric phase has the same origin as $\nu_m$ introduced here, it cannot be directly applied to the exceptional sphere here due to the two-fold degeneracy of the bands. As the spectrum is purely real outside and purely imaginary inside the sphere, the topological invariant defined above takes a nontrivial value. 

To reveal the relation between the PBC and OBC systems, we again adopt the transfer matrix method used in Eqs.~\eqref{kinetic-transfer-hal}$-$\eqref{transfer-matrix}. The hopping matrix is found to be  $J(\tilde{\mathbf{k}})=1/(2i)\Gamma_2-1/2\Gamma_3$, and the on-site matrix is 
$M(\tilde{\mathbf{k}})=\sin k_x\Gamma_1+M_{\tilde{\mathbf{k}}}\Gamma_3+\lambda\gamma_4$. According to the expression for the transfer matrix $T$ in Eq.~\eqref{transfer-matrix}, its determinant is calculated to be $|\det T|=1$, which is unimodular. We emphasize that different from the case of the kinetic non-Hermitian potentials, there is no structure of the transfer matrix, namely, it cannot be further reduced by block-diagonalization. The unimodular transfer matrix indicates that PBC and OBC systems are consistent with each other. It is expected that the conventional bulk-boundary correspondence is preserved in this system.

%%%%%%%%%%%%%%%%%%%%%%%%%%%%%%%%%%%%%%%%%%%%%%%%%%%%%%%%%
\begin{figure}[t]
	\centering
	\includegraphics[width=0.9\linewidth]{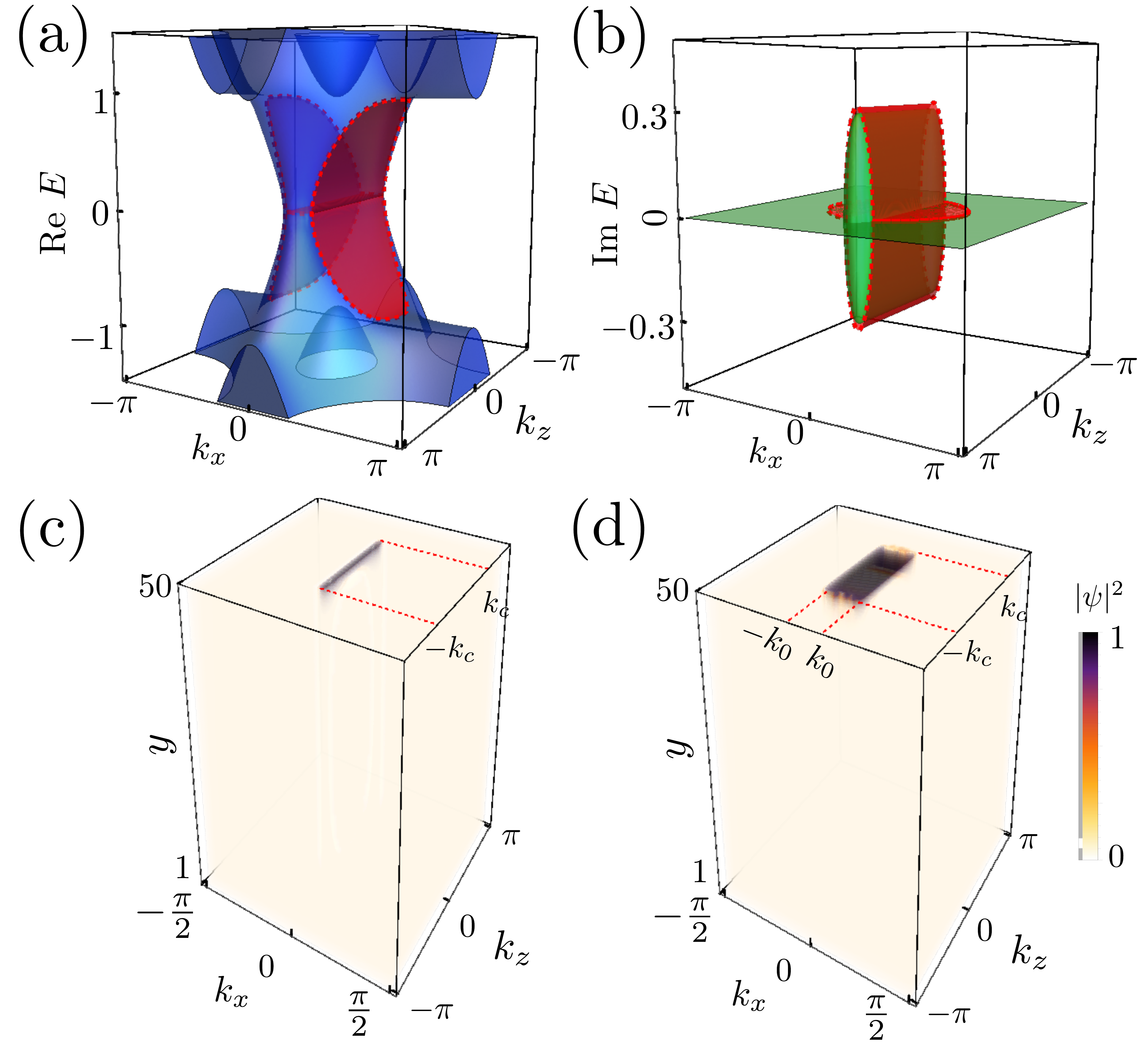}
	\caption{  
(a) Real and (b) imaginary parts of the OBC spectrum for the  $\PT$-symmetric Dirac semimetal with non-Hermitian anti-commuting potential. The parameters are $M=2.0$, $\lambda=0.3$, and $N_y=50$.
(c) Fermi arc surface states in Hermitian Dirac semimetals ($\lambda=0$), colored according to their wavefunction amplitude against $(k_x, k_z,y)$. The zero modes with $E=0$ form an arc (black curve) on the surface Brillouin zone.
(d) Fermi ribbon surface states in non-Hermitian Dirac semimetals ($\lambda = 0.3$), colored according to their wavefunction amplitude against $(k_x, k_z,y)$. The zero modes with $\re E=0$ form a two-dimensional ribbon (black region) on the surface Brillouin zone. }
	\label{FIG:anticommute}
\end{figure}
%%%%%%%%%%%%%%%%%%%%%%%%%%%%%%%%%%%%%%%%%%%%%%%%%%%%%%%%%

\subsection{$\mathbb{Z}_2$ monopole charge of the exceptional sphere}
With PBCs, outside the exceptional sphere in momentum space, the system is in the $\PT$-unbroken phase. For each $\mathbf{k}$ in this region, a set of eigenstates of $\mathcal{H}_\text{ant}(\mathbf{k})$
can be found satisfying 
\begin{equation}\label{realitycondition}
|\alpha(\mathbf{k})\rangle=\P\T|\alpha(\mathbf{k})\rangle,
\end{equation}
which is the same as the Hermitian case. With $(\P\T)^2=1$ the reality condition is imposed by $\PT$ symmetry on each band, and thus the eigenstates make a real Berry bundle in the $\PT$-unbroken region, similar to the Hermitian Dirac point case. In this section we show that the real $O(N)$ monopole charge, which is the same as for the Dirac point, can be obtained from the real Berry bundle in the $\PT$-unbroken region.

The existence of a real monopole charge indicates that if we enclose the gapless manifold, i.e., the exceptional sphere, by a sphere $S^2$ in the $\PT$-unbroken region, we cannot find a real and smooth gauge for the eigenvectors on the entire sphere~\cite{zhao_$pt$-symmetric_2017,fang_topological_2015}. We choose the radius of $S^2$ to be $k$, with $k>\lambda$. For simplicity, the Hamiltonian is scaled with $1/k$, so that the continuum model becomes $\mathcal{H}_\text{sphere}(\mathbf{k})/k = \mathcal{H}_\text{Dirac}(\mathbf{k})/k+\lambda_k \gamma_4$, with $\lambda_k=\lambda/k<1$. Correspondingly, the $S^2$ is scaled to a unit sphere. We need to analyze eigenvectors on the north and south hemispheres separated by an equator with stereographic coordinates. On each hemisphere the eigenvectors are real and smooth. On the equator parametrized by the azimuthal angle $\phi$, the eigenvectors of the valence band for the north hemisphere are $|-,1\rangle^N=(\cos\phi,-\sqrt{1-\lambda_k^2},\lambda_k+\sin\phi,0)^T$  and $|-,2\rangle^N=(\lambda_k-\sin\phi,0,\cos\phi,-\sqrt{1-\lambda_k^2})^T$, and for the south hemisphere, they are $|-,1\rangle^S=(\sqrt{1-\lambda_k^2},-\cos\phi,0,\lambda_k+\sin\phi)^T$  and $|-,2\rangle^S=(0,\lambda_k-\sin\phi,\sqrt{1-\lambda_k^2},-\cos\phi)^T$.
Here $T$ denotes the vector or matrix transposition. The real transition function $g_{SN}^\mathbb{R}\in O(2)$ on the equator defined by $|-,\alpha\rangle^S=[g_{SN}^\mathbb{R}]_{\alpha\beta}|-,\beta\rangle^N$ is obtained as 
\begin{equation}\label{transfunc}
g_{SN}^\mathbb{R}(\phi)=\begin{pmatrix}
\cos\phi & -\sin\phi \\ 
\sin\phi & \cos\phi
\end{pmatrix}.
\end{equation}
The winding number for the transition function defined as a map from $S^1$ to $O(2)$ is $+1$, which corresponds to a nontrivial element of $\pi_1[O(2)]=\mathbb{Z}_2$. Actually, the above expression for the real transition function is in the same form as that of the Dirac point, which can be seen by simply setting $\lambda=0$. Thus the exceptional sphere has the same real $\mathbb{Z}_2$ monopole charge as the Dirac point in Eq.~(\ref{Dirac_hal}). It is noteworthy that the $\mathbb{Z}_2$ charge adopted here is independent of $\nu_m$. The former reflects whether a smooth gauge for the wavefunctions can be found over the Brillouin zone, while the latter is explicitly associated with the exceptional sphere itself. In the following, we discuss the topological surface states and their relation with the bulk topological invariants.

\subsection{Fermi ribbon surface states}

Taking OBCs in the $y$ direction, the lattice model for the Dirac semimetal with the $\PT$-symmetric non-Hermitian anti-commuting potential reads
\begin{multline}\label{anti-commute-obc}
\mathcal{H}_\text{ant}(\tilde{\mathbf{k}})=
\frac{1}{2i}\Gamma_2\otimes(\widehat{S}-\widehat{S}^\dagger)-\frac{1}{2}\Gamma_3\otimes(\widehat{S}+\widehat{S}^\dagger)
\\ +\left(\sin k_x\Gamma_1+ M_{\tilde{\mathbf{k}}}\Gamma_3+\lambda\gamma_4\right)\otimes\mathbbm{1}_{N_y}.
\end{multline}
The OBC and PBC systems are consistent with each other as discussed previously with the transfer matrix method. This is reflected in the OBC system as the separation between $\PT$-broken and $\PT$-unbroken regions, similar to PBC systems. As shown in Figs.~\ref{FIG:anticommute}(a) and~\ref{FIG:anticommute}(b), the spectrum is purely real in the region $|E|>|\lambda|$ and purely imaginary in the region   $|E|<|\lambda|$, where $E$ is the energy of the original Hermitian system. 

To obtain the surface states, we adopt the ansatz $|\psi_{\tilde{\mathbf{k}}}\rangle=\sum_{i=1}^{N_y}\beta^i|\xi_{\tilde{\mathbf{k}}}\rangle\otimes|i\rangle$ for the right eigenvector of the surface states. Here $ |\xi_{\tilde{\mathbf{k}}}\rangle$ is a spinor, $i$ labels the latice sites along the $y$ direction, and $\beta$ is a scalar with $|\beta|<1$. Solving the Schr\"{o}dinger equation $\mathcal{H}_\text{ant}(\tilde{\mathbf{k}})|\psi_{\tilde{\mathbf{k}}}\rangle=E(\tilde{\mathbf{k}}) |\psi_{\tilde{\mathbf{k}}}\rangle$, it is found that $\beta=M-\cos k_x-\cos k_z$, and the surface states correspond to the positive eigenvalue $+1$ of $i\Gamma_3\Gamma_2$.
In the subspace spanned by the eigenstates $\{\Psi_\uparrow,\Psi_\downarrow\}$ of $i\Gamma_3\Gamma_2$ with eigenvalues $+1$, the effective surface Hamiltonian is found to be
\begin{equation}
\mathcal{H}_\text{b}(\tilde{\mathbf{k}})=\sin k_x\sigma_1+i\lambda \sigma_3,
\end{equation}
in the region of  $|\beta|=|M-\cos k_x-\cos k_z|<1$. The calculation details are shown in Appendix~\ref{boundary-anti-commute}.
The energy eigenvalues are $E_\text{b}(\tilde{\mathbf{k}})=\pm\sqrt{\sin^2k_x-\lambda^2}$, with the eigenstates $|\psi_\text{b}(\tilde{\mathbf{k}})\rangle = (i\lambda+E_\text{b}(\tilde{\mathbf{k}}),\sin k_x)^T$. 
As shown in Figs.~\ref{FIG:anticommute}(a) and~\ref{FIG:anticommute}(b), the analytically obtained surface states are plotted in red, which are in perfect agreement with  the numerical results shown in blue and green.

The topological surface states are found to be attached to the projected center of the exceptional spheres at $\pm k_c$ on the $k_z$ axis, which verifies that the conventional bulk-boundary correspondence is preserved. Remarkably, the zero-energy modes ($\re E=0$) now form a ribbon on the surface $k_xk_z$ Brillouin zone, which we call Fermi ribbon surface states. For comparison, in Fig.~\ref{FIG:anticommute}(c), we show the Fermi arc surface states in the Hermitian case, which connect the projection of Dirac points on the surface $k_xk_z$ surface Brillouin zone (black curve). After introducing the non-Hermiticity, 
as can be seen from the surface spectrum $E_\text{b}(\tilde{\mathbf{k}})$, for $-k_c<k_z<k_c$, the non-Hermiticity extends the zero-mode region from $k_x=0$ to $-k_0<k_x<k_0$ with $k_0=\arcsin\lambda$, resulting in a ribbon on the $k_xk_z$-plane. Thus, the Fermi arc is extended to Fermi ribbon surface states, as shown by the black region in Fig.~\ref{FIG:anticommute}(d) where the wavefunction profiles of the zero modes are plotted. It is clear that the Fermi ribbon surface states are localized inside the ribbon region at the boundary. Notably, the boundary of the ribbon consists of exceptional points, and inside the ribbon, pairs of boundary modes with opposite imaginary energies acquire balanced gain and loss due to non-Hermiticity.

\section{Summary and discussion}
\label{conclusion-and-discussion}

In summary, we systematically studied $\PT$-symmetric Dirac semimetals on periodic lattices perturbed by general non-Hermitian 
potentials. We found that in general there are only two different types of symmetry allowed non-Hermitian potentials, namely non-Hermitian kinetic potentials and non-Hermitian anti-commuting potentials. For both non-Hermitian potentials we investigated the band topology, the bulk exceptional points, and the surface states. 
Interestingly, we find that on a system with periodic boundary conditions, the non-Hermitian kinetic potential induces exceptional rings, while the
non-Hermitian anti-commuting potential leads to exceptional spheres. The latter are characterized by a $\mathbb{Z}_2$ monopole charge,
similar to the Hermitian case~\cite{zhao_$pt$-symmetric_2017}. With open boundary conditions, we find that 
the non-Hermitian kinetic potential gives rise to a non-Hermitian skin effect on both sides of the sample,
while the non-Hermitian anti-commuting potential generates Fermi ribbon surface states. These surface states
have vanishing real energy and exist in a two-dimensional region of the surface Brillouin zone that connects
the two projected exceptional spheres.  Moreover, the Fermi ribbon surface states
are bounded by one-dimensional lines of exceptional points, i.e., by exceptional rings.

The $\PT$-symmetric non-Hermitian Dirac semimetal studied in this work can be experimentally realized in metamaterials,
for example, in  acoustic metamaterials~\cite{valerio_huber_acoustic_Weyl_Nat_Phys_2019}, periodic electric circuits~\cite{ching_thomale_comm_phys_18}, 
or photonic lattices~\cite{guo_three_2017,Guo_observation_2019}.
In the latter systems radiation gain and loss are naturally present, leading to non-Hermitian potentials.
The type and strength of the non-Hermitian potential can be controlled by adjusting the different couplings between
each photonic mode and the environment~\cite{zhen_spawning_2015,zhou_observation_2018}. Notice that the creation of $\PT$-symmetric non-Hermitian potentials does not require perfect balanced gain and loss. In a totally passive system, these potentials can also emerge with a global loss offset in the passive background~\cite{feng_non-hermitian_2017}. 
It would be particularly interesting to realize the Fermi ribbon states and exceptional rings at the surface of photonic lattices, as these could potentially 
be used for applications, e.g., for surface sensing.   

Our work can be extended in a straightforward manner to other semimetals with $\PT$ symmetry~\cite{pt_cp_top_semi_metal_PRL_16}, for example, $\PT$-symmetric nodal-line
semimetals. Other directions for future work include  the study of non-Hermitian topological phase transitions~\cite{Chen_2019} and the derivation of a general classification of exceptional spheres based on not only fundamental~\cite{okugawa_topological_2019,kawabata_classification_2019,budich_symmetry-protected_2019} but also crystalline symmetries.

\section{Acknowledgments}
The authors thank Yuxin Zhao for in-depth discussions. The authors thank Song-Bo Zhang, Darshan Joshi, Yi Lu, and Satoshi Ikegaya for their valuable comments.

\appendix
\section*{Appendices} 

%========================================================================================
\section{Topological invariants of exceptional manifolds}
\label{jordan-normal-form}

In this appendix, we calculate the Jordan normal forms of the exceptional rings and spheres, and explicitly obtain the topological invariants associated with them that are defined by $\nu_m(S^1)$ and $\nu_m(S^0)$ in the main text. We show that this kind of topological invariants is closely related to the Jordan canonical form at exceptional points, and we propose a $\mathbb{Z}/N$ classification of non-Hermitian Hamiltonians with exceptional points.

\subsection{Exceptional ring}
The Hamiltonian for the exceptional rings in the continuum limit reads 
\begin{equation}
H_\text{ring}= k_x\Gamma_1+ k_y\Gamma_2+ k_z\Gamma_3+\lambda\gamma_2,
\end{equation}
with the energy eigenvalues of $E_{\text{ring}}= \pm \sqrt{k_x^2+ k_y^2+ k_z^2-\lambda^2 \pm 2i\lambda k_y}.$ The band crossing happens at $k_x^2+k_z^2=\lambda^2$ and $k_y=0$, which forms an excpetional ring on the $k_y=0$ plane. By parametrizing the ring as $(k_x,k_y,k_z)=\lambda(\cos\theta,0,\sin\theta)$ and after a unitary transformation by $U$, the Hamiltonian on the ring reads
\begin{multline}
H_{\text{exp-ring}} =\\\lambda\begin{pmatrix}
\sin\theta& 1+\cos\theta &0  &0  \\ 
-1+\cos\theta&-\sin\theta  & 0  &  0\\ 
0&0  &\sin\theta  & 1+\cos\theta  \\ 
0&0  &-1+\cos\theta  &-\sin\theta 
\end{pmatrix}.
\end{multline}
It turns out there are only two eigenvectors on the exceptional rings, 
\begin{equation}
\begin{aligned}
&|\psi_1\rangle=\begin{pmatrix}
0&
0 &
-1-\cos{\theta} &
\sin\theta
\end{pmatrix}^T ,\\ &|\psi_2\rangle=\begin{pmatrix}
-1-\cos{\theta} &
\sin\theta &
0 &
0
\end{pmatrix}^T.
\end{aligned}
\end{equation}
Because there are only two independent eigenvectors, it can be indicated that the largest Jordan block is two-dimensional. Besides the above two eigenvectors, there are two associated vectors with the Hamiltonian, satisfying $ H_{\text{exp-ring}}|\psi_1'\rangle=|\psi_1\rangle$ and $ H_{\text{exp-ring}}|\psi_2'\rangle=|\psi_2\rangle$. Together with eigenvectors, they form a matrix $P=(|\psi_1\rangle,|\psi_1'\rangle,|\psi_2\rangle,|\psi_2'\rangle)$, which reads
\begin{equation}
P=\begin{pmatrix}
0 & 0 & -1-\cos {\theta} &0 \\ 
0 & 0 & \sin{\theta} & -1\\ 
-1-\cos {\theta} &0& 0 & 0 \\ 
 \sin{\theta} & -1 & 0 & 0
\end{pmatrix}.
\end{equation}
Under the operation by $P^{-1} H_{\text{exp-ring}}P= J_{\text{ring}}$, the Jordan canonical form of the Hamiltonian on the exceptional ring can be obtained as
\begin{equation}
J_{\text{ring}}=\begin{pmatrix}
0 & 1 & 0 & 0 \\ 
0 & 0 & 0 & 0 \\ 
0 & 0 & 0 & 1 \\ 
0 & 0 & 0 & 0
\end{pmatrix}.
\end{equation}
The Jordan canonical of the exceptional ring Hamiltonian consists of two two-dimensional Jordan blocks. As we will show, the non-trivial Jordan block is associated with a topological invariant that is calculated below. 

As the energy spectra become complex, it is possible that they possess a topological structure, which essentially can be characterized by the topological invariant $\nu_m(S^1)$ defined in the main text. As shown in Fig.~\ref{FIG:exceptional-manifold} (b), we can enclose the exceptional ring (red) by a small circle $S^1$ (blue).
The small circle $S^1$ can be parameterized as $(\delta k_x,\delta k_y,\delta k_z)=(0,\rho\sin\gamma,\lambda-\rho\cos\gamma)$, then the eigenvalues for the conductance band on this path are
\begin{equation}
E_{S^1,\pm}= \sqrt{\rho^2-2\lambda\rho\cos\gamma\pm2i\lambda\rho\sin\gamma}.
\end{equation}
With $\rho \ll \lambda$, the eigenvalues can be approximated as $E_{S^1,\pm}= i \sqrt{2\lambda\rho} e^{\mp i\gamma/2}$. The topological invariant for $ E_{S^1,+}$ can be obtained as
\begin{equation}
\begin{aligned}
\nu(S^1)&=-\frac{1}{2\pi}\oint_{S^1} \nabla_{\mathbf{k}} \arg E_{S^1,+}(\mathbf{k})\\
&=-\frac{1}{2\pi} \int_{0}^{2\pi} d\gamma\partial_\gamma (-\gamma/2)= +\frac{1}{2}.
\end{aligned}
\end{equation}

\subsection{Exceptional sphere}
The Hamiltonian for the exceptional sphere in the continuum limit reads 
\begin{equation}
H_\text{sphere}=k_x\Gamma_1+ k_y\Gamma_2+ k_z\Gamma_3+\lambda\gamma_4,
\end{equation}
with the eigenvalues  $E_{\text{sphere},\pm}=\pm \sqrt{k_x^2+k_y^2+k_z^2-\lambda^2}$. The exceptional surface with zero eigenvalues can be parameterized by $(k_x,k_y,k_z)=$ $\lambda$($\sin \theta\cos\phi$, $\sin \theta\sin\phi$,$\cos\theta$), and after a unitary transformation by $U$, the Hamiltonian on the sphere is
\begin{equation}
H_{\text{exp-sphere}}=\lambda\begin{pmatrix}
\cos\theta & \sin\theta e^{-i\phi} & 0 & 1 \\ 
\sin\theta e^{i\phi} & -\cos\theta & -1 & 0 \\ 
0 & 1 & \cos\theta & \sin\theta e^{i\phi} \\ 
-1 & 0 & \sin\theta e^{-i\phi} & -\cos\theta
\end{pmatrix}.
\end{equation}
From this Hamiltonian, the eigenvectors on the exceptional surface are
\begin{equation}
\begin{aligned}
|\psi_1\rangle&=
\begin{pmatrix}
\cos\theta&
e^{i\phi} \sin\theta&
0 &
-1
\end{pmatrix}^T ,\\
|\psi_2\rangle&=
\begin{pmatrix}
e^{-i\phi} \sin\theta &
-\cos \theta &
1&
0
\end{pmatrix}^T,
\end{aligned}
\end{equation}
which span a two-dimensional eigenspace. Similar to the exceptional ring case, the largest Jordan block is two dimensional. Together with the two associated vectors, which satisfy $ H_{\text{exp-sphere}}|\psi_1'\rangle=|\psi_1\rangle$ and $ H_{\text{exp-sphere}}|\psi_2'\rangle=|\psi_2\rangle$, they form a matrix $P=(|\psi_1\rangle,|\psi_1'\rangle,|\psi_2\rangle,|\psi_2'\rangle)$, which reads 
\begin{equation}
P=\begin{pmatrix}
\cos\theta & 1 & e^{-i\phi}\sin\theta & 0 \\ 
e^{i\phi}\sin\theta & 0 & -\cos\theta & 1 \\ 
0 & 0 & 1 & 0 \\ 
-1 & 0 & 0 & 0
\end{pmatrix}.
\end{equation}
The Jordan canonical form of the Hamiltonian on the exceptional sphere can be obtained by performing $P$ onto the Hamiltonian, $J_{\text{sphere}}=P^{-1}H_{\text{exp-sphere}} P$, which results in
\begin{equation}
J_{\text{sphere}}=\begin{pmatrix}
0 & 1 & 0 & 0 \\ 
0 & 0 & 0 & 0 \\ 
0 & 0 & 0 & 1 \\ 
0 & 0 & 0 & 0
\end{pmatrix}.
\end{equation}

In the next section, we calculate the topological invariant associated with the exceptional sphere, as defined by $\nu_m(S^0)$ in the main text. The exceptional sphere can be enclosed by a zero-dimensional sphere $S^0$, as shown in Fig.~\ref{FIG:exceptional-manifold} (c) by two blue dots $\mathbf{k}_1$ and $\mathbf{k}_2$. The eigenvalue for the conduction bands for $\mathbf{k}_1$ is $E_1=\sqrt{k_1^2-\lambda^2}$ and for $\mathbf{k}_2$, it is $E_2=\sqrt{k_2^2-\lambda^2}$. The magnitudes of $\mathbf{k}_1$ and $\mathbf{k}_2$ satisfy $k_1<\lambda$ and $k_2>\lambda$. By the definition of our topological invariant in the main text, the topological charge can be computed as
\begin{equation}
\nu(S^0)=\frac{1}{2\pi} (\arg E_1-\arg E_2)=\frac{1}{2},
\end{equation}
which indicates that the exceptional sphere is topologically stable.

\subsection{$\mathbb{Z}/N$ classification of non-Hermitian Hamiltonians with exceptional points}\label{fractal}
The topological invariants defined for the exceptional ring and sphere have their roots in the Jordan canonical form. In this section, based on the general form of Jordan block, we generalize the definition of these topological invariants. We propose a $\mathbb{Z}/N$ type topological invariant for non-Hermitian Hamiltonians $H(\mathbf{k})$ with exceptional point of order $N$. The order of the exceptional point is defined by the dimension of the non-diagonal Jordan block of the Hamiltonian at the exceptional points. For order $N$ exceptional points at $\mathbf{k}_0$, the non-diagonal Jordan block is of the form,
\begin{equation}
J_N(\mathbf{k}_0)= \begin{pmatrix}
E_0 & 1 & 0 & \dots & 0 \\ 
0 & E_0 & 1 & \dots & 0 \\ 
\vdots & \vdots & \vdots & \ddots & \vdots \\ 
0 & 0 & 0 & \dots & 1 \\ 
0 & 0 & 0 & \dots & E_0
\end{pmatrix}_{N\times N},
\end{equation}
with the energy eigenvalue $E_0$ at the exceptional point. In the neighborhood ($|\delta \mathbf{k}|=\varepsilon$) of the exceptional points, we can adopt Newton-Puiseux series to investigate the bifurcation of eigenvalues, which is $ E=E_0+\sum_{i=1}^{+\infty}E_i \varepsilon^{i/N}$ \cite{demange_signatures_2012,seyranian2003multiparameter}. The eigenvalue to the leading order is 
\begin{equation}\label{series}
E=E_0+E_1 \varepsilon^{1/N}+o(\varepsilon^{1/N}),
\end{equation}
where
\begin{equation}\label{winding}
E_1^N=\sum_i \langle \mathbf{v}_0| \frac{\partial H(\mathbf{k})}{\partial k_i} |\mathbf{u}_0\rangle \frac{\partial k_i}{\partial \varepsilon}.
\end{equation}
Here $\langle \mathbf{v}_0|$ and $|\mathbf{u}_0\rangle$ are the left and right eigenvector satisfying 
\begin{equation}
\langle \mathbf{v}_0| H(\mathbf{k}_0)=\langle \mathbf{v}_0|E_0  \quad \text{and}\quad   H(\mathbf{k}_0) |\mathbf{u}_0\rangle=E_0 |\mathbf{u}_0\rangle.
\end{equation}
The right hand side of Eq.~(\ref{winding}) is complex, and can be expressed in the form of $|E_1^N| e^{i\theta(\mathbf{k})}$. Thus the eigenvalue to the leading order becomes,
\begin{equation}
E(\mathbf{k})=E_0+ |E_1| e^{i\theta(\mathbf{k})/N}.
\end{equation}
By encircling the exceptional points in $k$ space with a circle $S^1$,
a topological invariant can be defined as
\begin{equation}
v=\oint_{S^1} \frac{d \mathbf{k}}{2\pi i} \partial_\mathbf{k} \arg E(\mathbf{k}),
\end{equation}
which belongs to the group $\mathbb{Z}/N$.

\section{Effective surface Hamiltonian for the Fermi ribbon surface states}
\label{boundary-anti-commute}

In this appendix, we calculate the topological surface states for the case of exceptional sphere in the bulk. The PBC Hamiltonian in momentum space reads,
\begin{multline}
\mathcal{H}_\text{ant}(\mathbf{k})=\sin k_x\Gamma_1+\sin k_y\Gamma_2\\+(M-\cos k_x-\cos k_y-\cos k_z)\Gamma_3+\lambda \gamma_4.
\end{multline}
Taking OBCs in the $y$ direction with $N_y$ sites, the Hamiltonian in real space reads
 \begin{multline}
 \mathcal{H}_\text{ant}(\tilde{\mathbf{k}})=
 \frac{1}{2i}\Gamma_2\otimes(\widehat{S}-\widehat{S}^\dagger)-\frac{1}{2}\Gamma_3\otimes(\widehat{S}+\widehat{S}^\dagger)
 \\ +\left(\sin k_x\Gamma_1+ M_{\tilde{\mathbf{k}}}\Gamma_3+\lambda\gamma_4\right)\otimes\mathbbm{1}_{N_y},
 \end{multline}
with $M_{\tilde{\mathbf{k}}}=M-\cos k_x-\cos k_y$. Here the translational operators are
\begin{equation}
	\widehat{S}=
	\begin{pmatrix}
		0 & 0 & 0 & 0& \cdots & 0 \\ 
		1 & 0 & 0 & 0  & \cdots& 0 \\ 
		0 & 1 & 0 & 0 & \cdots & 0 \\ 
		0 & 0 & 1  & 0 & \cdots& 0 \\ 
		\vdots & \vdots & \vdots & \ddots & \ddots & \vdots \\ 
		0 & 0  & 0  & 0 & 1 & 0\\ 
	\end{pmatrix}, \widehat{S}^\dagger=
	\begin{pmatrix}
		0 & 1 & 0 & 0& \cdots & 0 \\ 
		0 & 0 & 1 & 0  & \cdots& 0 \\ 
		0 & 0 & 0 & 1 & \cdots & 0 \\ 
		0 & 0 & 0  & 0 & \ddots& 0 \\ 
		\vdots & \vdots & \vdots & \vdots & \ddots &1 \\ 
		0 & 0  & 0  & 0 & 0 & 0\\ 
	\end{pmatrix},
\end{equation}
with dimension $N_y$.

With the boundary conditions, the translational symmetry in the $y$ direction is broken, while in the $x$ and $z$ directions it is still preserved. Thus we adopt the ansatz for the surface states as $|\psi_{\tilde{\mathbf{k}}}\rangle=\sum_{i=1}^{N_y}\beta^i|\xi_{\tilde{\mathbf{k}}}\rangle\otimes|i\rangle$, with $ |\xi_{\tilde{\mathbf{k}}}\rangle$ a spinor, $i$ the lattice sites along the $y$ direction, and $\beta$ a scalar with $|\beta|<1$. 

Solving the Schr\"{o}dinger equation $\mathcal{H}_\text{ant}(\tilde{\mathbf{k}})|\psi_{\tilde{\mathbf{k}}}\rangle=E(\tilde{\mathbf{k}}) |\psi_{\tilde{\mathbf{k}}}\rangle$, 
we find it gives two constraints, 
\begin{multline}\label{Dirac-mass-relation1}
\big[ \sin k_x\Gamma_1+\frac{1}{2i}(\beta-\beta^{-1})\Gamma_2+(M_{\tilde{\mathbf{k}}}-\frac{1}{2} (\beta+\beta^{-1}))\Gamma_3\\+\lambda\gamma_4 \big] |\xi_{\tilde{\mathbf{k}}}\rangle=E(\tilde{\mathbf{k}})|\xi_{\tilde{\mathbf{k}}}\rangle,
\end{multline}
and
\begin{multline}\label{Dirac-mass-relation2}
\big[\sin k_x\Gamma_1+\frac{1}{2i}\beta\Gamma_2+(M_{\tilde{\mathbf{k}}} -\frac{1}{2} \beta)\Gamma_{3}\\
+\lambda \gamma_{4}\big] |\xi_{\tilde{\mathbf{k}}}\rangle=E(\tilde{\mathbf{k}})|\xi_{\tilde{\mathbf{k}}}\rangle.
\end{multline}
From the difference between the above two equations, it can be obtained that
\begin{equation}\label{Dirac-mass-project1}
i\Gamma_{3}\Gamma_2|\xi_{\tilde{\mathbf{k}}}\rangle =|\xi_{\tilde{\mathbf{k}}}\rangle,
\end{equation}
with $\beta=M-\cos k_x-\cos k_z$. 

Clearly, the boundary states correspond to the positive eigenvalue $+1$ of the operator $i\Gamma_3\Gamma_2$. We choose the eigenvectors with eigenvalue $+1$ of $i\Gamma_3\Gamma_2$ as
\begin{align}
|\Psi_\uparrow\rangle=\frac{1}{\sqrt{2}}\begin{pmatrix}
-i&
0 &
0 &
1
\end{pmatrix}^T,\\
|\Psi_\downarrow\rangle=\frac{1}{\sqrt{2}}\begin{pmatrix}
0&
-i &
1 &
0
\end{pmatrix}^T.
\end{align}
The effective surface Hamiltonian can be obtained by the projection to the subspace spanned by the eigenstates $\{|\Psi_\uparrow\rangle,|\Psi_\downarrow\rangle\}$, 
\begin{equation}
\begin{aligned}
\mathcal{H}_\text{b}(\tilde{\mathbf{k}})&=\begin{pmatrix}
\langle \Psi_\uparrow|\mathcal{H}_\text{ant}(\mathbf{k}) |\Psi_\uparrow\rangle & \langle \Psi_\uparrow|\mathcal{H}_\text{ant}(\mathbf{k}) |\Psi_\downarrow\rangle \\ 
\langle \Psi_\downarrow|\mathcal{H}_\text{ant}(\mathbf{k}) |\Psi_\uparrow\rangle& \langle \Psi_\downarrow|\mathcal{H}_\text{ant}(\mathbf{k}) |\Psi_\downarrow\rangle
\end{pmatrix} \\
&=\sin k_x\sigma_1+i\lambda \sigma_3,
\end{aligned}
\end{equation}
in the region of  $|\beta|=|M-\cos k_x-\cos k_z|<1$.

\bibliographystyle{apsrev4-1}

\bibliography{NonHermi_Dirac_Ref}

\end{document}